\documentclass[aps,prd,a4paper,twocolumn,amsmath,showpacs,superscriptaddress,nofootinbib,preprintnumbers]{revtex4-1}

\usepackage{verbatim}
\usepackage[T1]{fontenc}
\usepackage[utf8]{inputenc}
\usepackage[american]{babel}
\usepackage{epsfig}
\usepackage{graphicx}
\usepackage{booktabs}
\usepackage{multirow}
\usepackage{dcolumn}
\usepackage{amsmath}
\usepackage{mathtools}
\usepackage{amsfonts}
\usepackage{amssymb}
\usepackage{epstopdf}
\usepackage{bm}
\usepackage{siunitx}
\usepackage{braket}
\usepackage{enumitem}
\usepackage{soul}
\usepackage[table]{xcolor}
\usepackage{color}
\usepackage{transparent}
\usepackage{pifont}

\definecolor{navyblue}{rgb}{0.0, 0.0, 0.5}
\definecolor{royalblue}{rgb}{0.25, 0.41, 0.88}
\definecolor{cadmiumgreen}{rgb}{0.0, 0.42, 0.24}
\definecolor{blue-violet}{rgb}{0.54, 0.17, 0.89}
\definecolor{darkviolet}{rgb}{0.58, 0.0, 0.83}
\definecolor{orange(colorwheel)}{rgb}{1.0, 0.5, 0.0}

\usepackage{hyperref}
\hypersetup{
    colorlinks=true, 
    linkcolor=royalblue, 
    citecolor=magenta}

\newcommand\ee{\end{equation}}
\newcommand\be{\begin{equation}}
\newcommand\eea{\end{eqnarray}}
\newcommand\bea{\begin{eqnarray}}

\usepackage{booktabs}
\usepackage{multirow}
\usepackage{dcolumn}
\usepackage{colortbl}

\definecolor{magenta(process)}{rgb}{1.0, 0.0, 0.56}

\definecolor{darkspringgreen}{rgb}{0.09, 0.45, 0.27}

\definecolor{royalblue(web)}{rgb}{0.25, 0.41, 0.88}

\begin{document}

\title{First cosmological constraints combining Planck
with the recent gravitational-wave standard siren measurement of the Hubble constant}  

\author{Eleonora Di Valentino}
\email{eleonora.divalentino@manchester.ac.uk}
\affiliation{Jodrell Bank Center for Astrophysics, School of Physics and Astronomy, University of Manchester, Oxford Road, Manchester, M13 9PL, UK}

\author{Alessandro Melchiorri}
\email{alessandro.melchiorri@roma1.infn.it}
\affiliation{Physics Department and INFN, Universit\`a di Roma ``La Sapienza'', Ple Aldo Moro 2, 00185, Rome, Italy} 

\date{\today}

\preprint{}
\begin{abstract}
The recent observations of gravitational-wave and electromagnetic emission produced by the merger of the binary neutron-star system GW170817 have opened the possibility of using standard sirens to constrain the value of the Hubble constant. While the reported bound of $H_0=70_{-8}^{+12}$ at $68 \%$ C.L. is significantly weaker than those recently derived by observations of Cepheid variables, it does not require any form of cosmic ‘distance ladder’ and can be considered as complementary and, in principle, more conservative. Here we combine, for the first time, the new measurement with the Planck Cosmic Microwave Background observations in a $12$ parameters extended $\Lambda$CDM scenario, where the Hubble constant is weakly constrained from CMB data alone and bound to a low value $H_0=55^{+7}_{-20}$ km/s/Mpc at $68 \%$ C.L. We point out that the non-Gaussian shape of the GW170817 bound makes lower values of the Hubble constant in worst agreement with observations than what expected from a Gaussian form. The inclusion of the new GW170817 Hubble constant measurement therefore significantly reduces the allowed parameter space, improving the cosmological bounds on several parameters as the neutrino mass, curvature and the dark energy equation of state. 
\end{abstract}

\maketitle

\section{Introduction}

The recent observations of gravitational-wave and electromagnetic emission produced by the merger of the binary neutron-star system GW170817 \cite{GW170718} have opened the possibility of using standard sirens (see e.g. \cite{nissanke}) to constrain the value of the Hubble constant. Indeed, in \cite{natureH0} a constraint of $H_0=70_{-8}^{+12}$ km/s/Mpc at $68 \%$ C.L. has been reported.
While the obtained constraints are significantly weaker than those derived from observations of Cepheid variables, they do not require any form of cosmic ‘distance ladder’ and can be considered, in principle, as more conservative \cite{natureH0}. This point is particularly relevant since the current constraints based on luminosity distances from \cite{R16} report a value of $H_0=73.24\pm 1.74$ km/s/Mpc at $68 \%$ C.L. that is in tension at more than $3$ standard deviations with the result derived 
from observations of CMB anisotropies from the Planck experiment \cite{planck2015} (see e.g. \cite{freedmann}). Assuming a standard cosmological scenario based on a cosmological constant, the recent analysis of \cite{plancknewtau} gives indeed $H_0=66.93\pm0.62$ km/s/Mpc at $68 \%$ C.L..

Clearly, the recent GW170817 measurement cannot discriminate between these two values and at least more than $25$ additional observations of standard sirens are needed for reaching an uncertainty on $H_0$ useful to scrutinize the tension (see Figure 2 in \cite{nissanke}). However, while the $H_0$ constraint from \cite{R16} could be affected by systematics, the $H_0$ determination from Planck is completely model-dependent and fully relies on the assumption of $\Lambda$CDM. For example, just assuming a dark energy equation of state $w$ in its simplest form (constant with redshift) and/or a non-flat universe introduces a geometrical degeneracy that makes the Hubble constant value from Planck as practically unbounded.  

It is therefore timely to investigate what is the impact of the new GW170817 $H_0$ measurement on the Planck constraints in an extended parameter space in which the Planck data alone is unable to strongly constrain the Hubble constant.

While the accuracy of the GW170817 determination is apparently relatively poor, we should also point out the fact that the $H_0$ posterior distribution presented in Figure 1 of \cite{natureH0} is strongly non-Gaussian, with a lower limit at $95 \%$ c.l. on the Hubble constant of $H_0 > 58$ km/s/Mpc (instead of $H_0 > 54$ km/s/Mpc in case of Gaussianity). This non-Gaussianity, due essentially to the unknown inclination plane of the binary orbit, must be taken into account when performing a combined analysis with CMB data.

Our paper is structured as follows: in the next section we describe our analysis method, in section III we illustrate our results and we conclude in Section IV.

\section{Method}

As we mentioned in the introduction, CMB is essentially unable to strongly constrain the Hubble constant once we also consider  parameters as the curvature of the universe or the dark energy equation of state. All these parameters can indeed affect the angular diameter distance of the last scattering surface $D_A$. Since the CMB is mostly sensitive to $D_A$, any combination of $H_0$, dark energy parameters and/or $\Omega_k$ that gives the same value for $D_A$ (and that preserves the sound horizon at recombination) will provide a nearly identical fit to the CMB data (see e.g. \cite{efstbond}).

We therefore work in an extended $12$ parameter space considering the usual $6$ parameters of the standard $\Lambda$CDM model but including also additional $6$ parameters that are degenerate or correlated with the Hubble constant when considering the CMB anisotropy angular spectra.

Of course considering more parameters at the same time results in weaker constraints, however it also properly takes statistically into account our observational ignorance about their values. Since, for example, we have no fundamental reason to believe that the Universe must be flat and/or that the dark energy component can be fully described by a cosmological constant it is in our opinion reasonable to consider these extensions. 

For the $6$, $\Lambda$CDM, parameters we consider the baryon and cold dark matter physical energy densities $\Omega_bh^2$ and $\Omega_{cdm}h^2$, the amplitude and the spectral index of primordial inflationary perturbations $A_S$ and $n_S$, the sound horizon angular scale at the last scattering surface $\theta_s$ and the reionization optical depth $\tau$.
To these six parameters we add: the absolute neutrino mass scale, $\Sigma m_{\nu}$, the running of the spectral index $n_S'=d n_S /dlnk$, the neutrino effective number $N_{eff}$, and the energy density in curvature $\Omega_k$. 
Moreover we parametrize the dark energy component using the Chevalier-Polarski-Linder parametrization:

\begin{equation}
w(a)=w_0+(1-a)w_a
\end{equation}

where $w_0$ is the value of the equation of state today while $w_a$ measures its time variation. In what follows we therefore let vary also $w_0$ and $w_a$ for a total of $12$ cosmological parameters varied {\it at the same time.}
following the approach used in \cite{paper, papero, paper1, paper2}.

On each parameter we assume flat priors as reported in Table~\ref{priors}. Please note that for the Hubble constant we choose a range in between $20<H_0<100$ km/s/Mpc, since we consider values out of this range as unphysical.

Our parameters constraints are obtained by firstly using the temperature and polarization CMB angular power spectra released by Planck 2015 \cite{Aghanim:2015xee}. This dataset includes both temperature and polarization anisotropies for the small angular-scale measured by the Planck High Frequency Instrument (HFI) experiment and for the large angular-scale measured by the Planck Low Frequency Instrument (LFI). In the following we refer to it as "Planck". 

We then include the GW prior on the Hubble constant. Since the posterior on $H_0$ is strongly non-gaussian, we use as GW prior an interpolation that can adequately reproduce the results in Figure 1 of \cite{natureH0}. In the following we refer to this prior as "GW170817".

The constraints are derived using the most updated version of the publicly available Monte Carlo Markov Chain package \texttt{cosmomc} \cite{Lewis:2002ah}, based on the Gelman and Rubin convergence diagnostic and that fully supports the Planck data release 2015 Likelihood Code \cite{Aghanim:2015xee} (see \url{http://cosmologist.info/cosmomc/}).

Foreground parameters are also varied following the procedure described in \cite{Aghanim:2015xee} and \cite{planck2015}.

\begin{table}
\begin{center}
\begin{tabular}{c|c}
Parameter                    & Prior\\
\hline
$\Omega_{\rm b} h^2$         & $[0.005,0.1]$\\
$\Omega_{\rm c} h^2$       & $[0.001,0.99]$\\
$\tau$                       & $[0.01,0.8]$\\
$n_s$                        & $[0.8, 1.2]$\\
$\log[10^{10}A_{s}]$         & $[2,4]$\\
$\theta_{\rm s}$             & $[0.5,10]$\\ 
$\sum m_\nu$ (eV)               & $[0,5]$\\
$w_0$ & [-3,0.3]\\
$w_a$ & [-2,2]\\
$N_{\rm eff}$ & [0.05,10]\\ 
$\frac{dn_s}{d\ln k}$ & [-1,1]\\
$\Omega_k$ & [-0.3,0.3] \\
\end{tabular}
\end{center}
\caption{External flat priors on the cosmological parameters assumed in this paper.}
\label{priors}
\end{table}

\section{Results}

\begin{table*}
\begin{center}
\begin{tabular}{c|cc}
Parameter                    & Planck & Planck+GW170817\\
\hline
\hspace{1mm}\\

$\Omega_{\rm b} h^2$         & $0.02231 \pm 0.00028$
& $0.02232 \pm 0.00028$\\
$\Omega_{\rm cdm} h^2$       & $0.1197 \pm 0.0035$
& $0.1195 \pm 0.0034$\\
$\tau$                       & $0.054^{+0.020}_{-0.024}$
& $0.058^{+0.020}_{-0.023}$\\
$n_s$                        & $0.968 \pm 0.012$
& $0.967 \pm 0.012$\\
$\log[10^{10}A_{s}]$         & $3.039^{+0.041}_{-0.050}$
& $3.050^{+0.041}_{-0.046}$\\
$\theta_{\rm s}$             & $1.04061 \pm 0.00051$
& $1.04069 \pm 0.00050$\\ 
$N_{\rm eff}$ & $3.11 \pm 0.25$
&$3.09 \pm 0.25$\\ 
$\frac{dn_s}{d\ln k}$ & $0.0038 \pm 0.0087$
& $0.0024 \pm 0.0086$\\

\hspace{1mm}\\
\hline
\hline
\hspace{1mm}\\

$\sum m_\nu$ (eV)            & $<1.11$
& $<0.77$\\
$w_0$ & unconstrained
&$-2.10^{+0.30}_{-0.84}$\\
$w_a$ & $-0.2^{+0.7}_{-1.7}$
& $<0.491$\\
$\Omega_k$ & $-0.068^{+0.058}_{-0.024}$ 
&$-0.025^{+0.013}_{-0.010}$\\
$H_0$ [km/s/mpc] & $54^{+7.0}_{-20}$
& $70.2^{+5.0}_{-9.8}$\\
$\sigma_8$ & $0.738^{+0.087}_{-0.16}$
& $0.893^{+0.066}_{-0.089}$\\
\hspace{1mm}\\

\hline
\end{tabular}
\end{center}
\caption{Constraints at $68$ \% c.l. on cosmological parameters from Planck 2015 before and after the inclusion of the GW170817 prior on the Hubble constant. The parameters below the double line are those mostly affected by the inclusion of the GW170817 prior. The total neutrino mass upper limits $\sum m_\nu$ are at 95\% C.L..}
\label{constraints}
\end{table*}

\begin{figure*}
\centering
\begin{tabular}{cc}
\includegraphics[width=0.9\columnwidth]{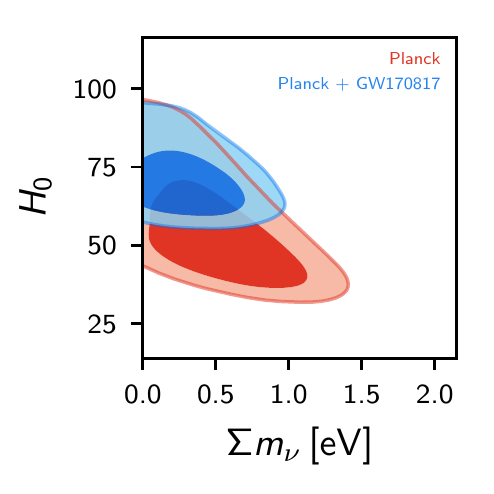} & \includegraphics[width=0.9\columnwidth]{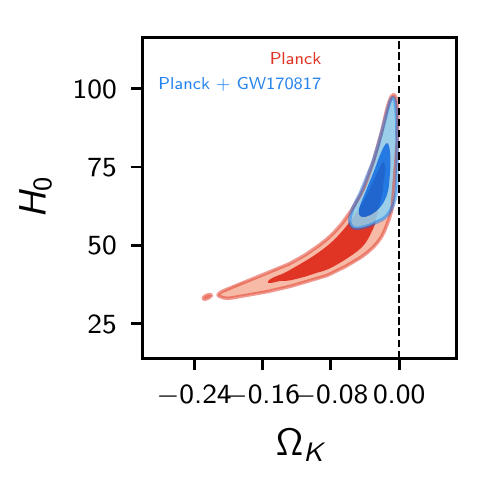}  \\
\includegraphics[width=0.9\columnwidth]{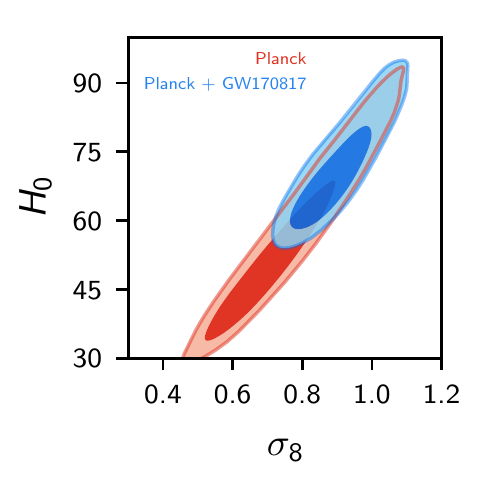}  &
\includegraphics[width=0.9\columnwidth]{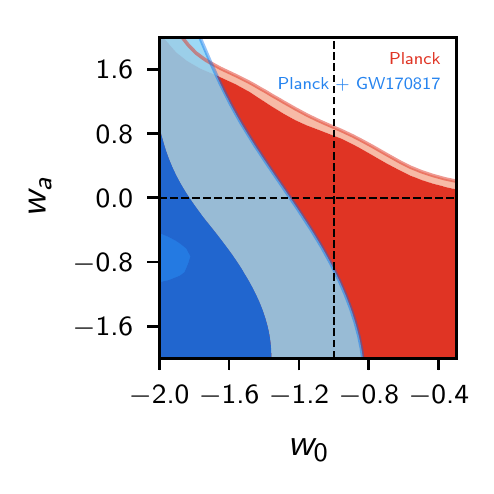} \\
\end{tabular}
\caption{68.3\% and 95.4\% confidence level constraints on a $12$ parameter extended space for the Planck and Planck+GW170817 data
in the $\Sigma m_{\nu}$ vs $H_0$, $\Omega_k$ vs $H_0$, $\sigma_8$ vs $H_0$, and $w_0$ vs $w_a$ planes.}
\label{fig1}
\end{figure*}

In Table~\ref{constraints} we report the parameter constraints at $68 \%$ C.L. from the Planck and Planck+GW170817 analyses.

If we first focus our attention on the bounds on $H_0$ we immediately see that in our $12$ parameters theoretical framework the Planck data seems to prefer quite low values for the Hubble constant ($H_0\sim 50$ km/s/Mpc). This preference is due to a parameter volume effect and is driven by degeneracies between $\Omega_k$, $w_0$, and $w_a$. In practice, a low $H_0$ model {\it per se} does not provide a better fit to the Planck data but since we have "more models" with low $H_0$ that give an equally good fit to CMB data the posterior distribution for $H_0$ is globally shifted towards lower values. This introduces an apparent small tension with the GW170817 prior that disfavors low $H_0$ models and makes its inclusion clearly significant. When the GW170817 prior is included, a large portion of models with low $H_0$ values is excluded and we get the constraint $H_0=70.2_{-9.8}^{+5.0}$ at $68 \%$ C.L.. The best fit $\chi^2$ before and after the inclusion of GW170817 remains essentially the same, clearly indicating that the tension on $H_0$ is mainly a volume parameter effect and that this prior can be safely combined with the Planck data.

From Table~\ref{constraints} we can identify $5$ parameters that are mostly degenerate with $H_0$ and that are better constrained when the GW170817 prior is included. These parameters are: the curvature of the universe $\Omega_k$, the neutrino absolute mass scale $\Sigma m_{\nu}$, the amplitude of r.m.s. matter density fluctuations $\sigma_8$, and the two dark energy parameters $w_0$ and $w_a$. This is clear also from Figure~\ref{fig1} where we show the 2D posteriors for the Planck and Planck+GW170817 datasets in the $\Sigma m_{\nu}$ vs $H_0$, $\Omega_k$ vs $H_0$, $\sigma_8$ vs $H_0$, and $w_0$ vs $w_a$ planes.

As we can see, despite the GW170817 prior being rather weak respect to other recent $H_0$ determinations, it significantly reduces the parameter space.
While the constraints on the baryon and cold dark matter densities, on the optical depth $\tau$ and on inflationary parameters are essentially left as unaltered by the inclusion of the GW170817 prior, we see that the constraints on the total neutrino mass are significantly stronger. Planck+GW170817 gives indeed an upper limit on the neutrino mass of $\sum m_\nu< 0.77 eV$ at $95\%$ C.L. that is about $\sim 30 \%$ stronger then the upper limit obtained from the Planck data alone. This is clear on the Top Left panel of Figure~\ref{fig1} where a degeneracy line between the neutrino mass and $H_0$ is evident. Higher values  for the neutrino mass are allowed for smaller values of $H_0$ that are at odds with the GW170817 prior.

A similar, strong, improvement is present in the case of curvature. As we can see from the Top Right panel of Figure~\ref{fig1} a quite significant number of models with large positive curvature is compatible with the Planck data for low ($<60$ km/s/Mpc) $H_0$ values. This portion of parameter space is excluded by the GW170817 prior and the constraints are improved by more than a factor $2$. It is interesting to notice that both the Planck and the Planck+GW170817 datasets prefers a closed universe at about 
$95 \%$ C.L.. We comment more about this point in the conclusions.

Geometrical degeneracy propagates on all parameters and affects also quantities like $\sigma_8$ that are not directly related to it. As we can see from Table~\ref{constraints} and from the Bottom Left panel of Figure~\ref{fig1}, the constraints on $\sigma_8$ from Planck are significantly improved when the GW170817 $H_0$ prior is included, ruling out a large region of models with low $\sigma_8$ and low $H_0$.

In the Bottom Right panel of Figure~\ref{fig1} we plot the 2D constraints on the $w_0$-$w_a$ plane with and without the GW170817 prior. As we can see, while the constraints are weak in both cases, when the GW170817 prior is included a whole class of models with $w_a>0$ and $w_0>-1$ appears to be in disagreement with the data. There is clearly some parameter volume effect: 
in practice, more models with $w_a<0$ and $w_0<-1$ provide a good fit to the data and this shifts the posterior distribution in this region. One should be therefore careful in concluding from the plot that a cosmological constant is excluded at $95 \%$ C.L. from the Planck+GW170817 dataset (see the dashed lines in the Figure). However clearly the inclusion of the GW170817 prior provides an upper limit on $w_0$ that was absent from the Planck data alone.

\section{Conclusions}

In this brief letter we have combined the recent standard siren estimate of the Hubble constant of $H_0=70_{-8}^{+12}$ km/s/Mpc at $68 \%$ C.L. of \cite{natureH0}
with the Planck CMB dataset to quantify the improvement in the constraints in the case of an extended $12$ parameters model, in which the Hubble constant is weakly constrained from CMB data alone.

We have found that including variations in $\Omega_k$, $w_0$, $w_a$ and in the total neutrino mass enlarges significantly the CMB bounds on $H_0$, making them less stringent than the GW170817 constraint and, perhaps more importantly, shifting them towards lower values of $H_0$, in slight tension with the GW170817 bound.
The inclusion of the GW170817 prior therefore improves significantly the Planck constraints on several parameters, most notably on curvature, neutrino mass, $\sigma_8$, and on the dark energy equation of state.

While these constraints should be regarded as conservative given the broad range of $H_0$ values allowed by the GW170817 prior, some tension with the standard $\Lambda$CDM model are present. In particular, a positive curved universe appears preferred at $95 \%$ C.L.. Also a phantom like dark energy equation of state seems preferred from the analysis.

The preference for positive curvature is already present in the Planck dataset alone and is probably connected to small anomalies in the Planck data (see e.g. discussion in \cite{paper,Addison:2015wyg,elldivide2,elldivide3}).

The indication for a phantom-like dark energy equation of state is driven instead by the GW170817 prior. In our $12$ parameters scenario, models with a lower value of the Hubble constant ($H_0<55$ km/s/Mpc) and that provide a good fit to the Planck data are in the $w_a>0$, $w_0>-1$ sector. The inclusion of the GW170817 prior exclude these models, giving rise to a preference for $w_0<-1$ models.

Before concluding it is important to notice that there are several external cosmological datasets as Baryon Acoustic Oscillations (BAO) \cite{beutler2011,ross2014,anderson2014}, cosmic shear data \cite{kids,deswl1,deswl2} and supernovae type Ia luminosity distance from the JLA catalog \cite{JLA}, just to name a few, that can provide much stronger constraints on the parameters considered here than the Planck+GW170817 case. Howewer some tension between these datasets exist. We note in particular that any indication for curvature disappears when a Planck+BAO dataset is considered while it will still be allowed when considering Planck+JLA
(see Table IV of ~\cite{paper1}).

The cosmological bounds presented here should be therefore considered as complementary and conservative. Future observations of standard sirens in the next years will certainly improve current estimates on $H_0$ and possibly shed light on the several tensions present between cosmological data.

\acknowledgments 
 
EDV acknowledges support from the European Research Council in the form of a Consolidator Grant with number 681431. AM thanks the University of Manchester and the Jodrell Bank Center for Astrophysics for hospitality.

\end{document}